\documentclass[useAMS,usenatbib,usegraphicx,english]{mn2e}
\usepackage{journals}
\usepackage{times}

\newcommand{\ms}{M$_{\odot}$}

\title{Radial migration in a bar-dominated disk galaxy I: Impact on chemical evolution}

\author[M. Kubryk, N. Prantzos and E. Athanassoula]
{M. Kubryk$^1$, N. Prantzos$^1$, E. Athanassoula$^2$ \\
$^1$UPMC-CNRS, UMR7095,Institut d'Astrophysique de Paris, 75014 Paris, France \\
$^2$Aix Marseille Universit\'e, CNRS, LAM (Laboratoire d’Astrophysique de Marseille) UMR 7326, 13388, Marseille, France\\
}

\date{}

\begin{document}

\maketitle

\begin{abstract}
We study radial migration and chemical evolution in a bar-dominated
disk galaxy, by analyzing the results 
of a fully self-consistent, high resolution N-body+SPH simulation. We find different behaviours for gas and star particles.
Gas within corotation is driven in the central regions by the bar, where it forms a
pseudo-bulge (disky-bulge), but it undergoes negligible radial displacement outside
the bar region. Stars undergo substantial radial migration at all
times, caused first by transient spiral arms and later by the bar.
Despite the important
amount of radial migration occurring in our model, its impact
on the chemical properties is limited. The reason is
the relatively flat abundance profile, due
to  the rapid early evolution of the whole disk. 
We show that the implications
of radial migration on chemical evolution can be studied to a good
accuracy by post-processing the results of the N-body+SPH calculation
with a simple chemical evolution model having detailed chemistry and
a parametrized description of radial migration. We find that radial
migration impacts on chemical evolution 
both directly (by 
moving  around the long-lived agents of
nucleosynthesis, like e.g. SNIa or AGB stars, and thus altering the  
abundance profiles of the gas) and indirectly 
(by moving around the long-lived tracers of chemical evolution and thus affecting
stellar metallicity profiles, local age-metallicity relations and  metallicity distributions
of stars, etc.). 
\end{abstract}


\section{Introduction}

In the past decade, observational and theoretical studies suggested that
radial migration of stars may play an important role in shaping the
properties of galactic disks. Already in the 1990s, it was realized that
the observed dispersion in the age-metallicity relation of the solar
neighborhood \citep{Edvardsson+93} was apparently
too large to be explained solely
by orbital diffusion due to epicyclic motions (i.e. by stars born in 
the inner Galaxy, in regions of metallicity higher than the local one). Moreover,
\cite{Wielen+96} argued that the Sun should have originated $\sim$2
kpc inwards from its current Galactocentric radius, in order to explain
its high metallicity with respect to the one of the local ISM and of nearby stars; that value was close to the maximum radial epicyclic excursion 
$\Delta R\sim \sqrt{2} \sigma_R /\kappa \sim$ 2 kpc (in view of the observed local values of radial velocity dispersion $\sigma_R \sim$50 km/s \citep{Holmberg2009}
and epicyclic frequency  $\kappa \sim$37 km/s/kpc \citep{BT08}). 
Since then, the solar metallicity has been revised downwards \citep{Asplund2009} and it is  compatible with the one of nearby young stars and of the local ISM; still, it is unclear whether observations require important radial mixing, i.e. beyond the one implied by epicyclic motions; see e.g. \cite{Nieva2012} vs \cite{Haywood2012}).

\cite{SellwoodBinney2002} (hereafter SB02) showed that,  in the
presence of recurring transient spirals, stars in a galactic disk could 
undergo radial displacements much larger than envisioned before: stars 
found at corotation with a spiral arm may  be scattered to different galactocentric radii (inwards or outwards,)  a process which preserves
overall angular momentum distribution and does not contribute to the radial heating of the stellar disk.
This development paved the way for a large number of theoretical studies on radial migration, both with numerical N-body codes and with semi-analytical models.

\cite{Lepine2003} considered a toy-model disk with corotation at a fixed galactocentric 
radius, removing stars locally and ''kicking" them inwards and outwards. They find that
in the case of the Milky Way disk, the abundance profile (assumed to be initially exponential) 
flattens in the 8-10 kpc region, but current data are inconclusive in that respect \citep{Luck2011}. 
Analyzing N-body+SPH simulations, \cite{Roskar2008a} presented a systematic investigation
of the implications of radial migration ({\it \`a la} SB02) for the chemical evolution of galactic disks; 
they discussed dispersion in the age-metallicity relation,
broadening of the metallicity distribution, flatenning of observed past abundance
profiles and flatenning of the observed past star formation history. Some of those
effects were analyzed in more detail 
with a simple toy model by 
\cite{Prantzos09}, who  found that the SBO2 mechanism produces
an age-dependent dispersion in the age-metallicity relation (because young stars
have insufficient time to migrate from far-away regions with different metallicities)
and showed how the tails of the local metallicity distribution can be affected
by that process.

\cite{SB09}
coupled a full chemical evolution code with a parametrized prescription of
radial migration, distinguishing epicyclic motions (''blurring")
from migration (''churning") due to
transient spirals. 
They found excellent agreement between the results
of their model and observations of the solar neighborhood and they suggested
that radial mixing could also explain the formation of the Galaxy's thick disk,
by bringing to the solar neighborhood a kinematically ''hot" stellar population
from the inner disk. That possibility was subsequently investigated
with N-body models, but controversial results are obtained up to now
(compare e.g. \cite{Loebman2011} to \cite{Minchev2012a}). The issue is still under debate and 
it is unclear whether the Galaxy's thick disk is due to secular evolution or to
the effects of past mergers e.g. \citep{Bournaud2009,Brook2012,Forbes2012,Steinmetz2012,Bekki2011}, 
while \cite{Bovy2012} suggest that the thick disk is not a distinct
component of the Milky Way.

\cite{minchevfamaey2010} suggested a different mechanism for radial migration
than transient recurring spirals, namely resonance overlap of the bar and spiral structure; this strongly nonlinear coupling leads to a more efficient
redistribution of angular momentum in the disk  and produces a stellar velocity dispersion increasing with time,  in broad agreement  with local observations. 
This bar-spiral coupling was studied in detail by \cite{Shevchenko2011} and  \cite{Brunetti2011}; the latter
 found that the extent of radial migration depends also
on the kinematic state of the disk, being reduced in the case of 
kinematically hot disks.
The radial motion of stars in disks was analyzed with N-body+SPH models for both non-barred \citep{Grand2012a} and barred \citep{Grand2012b} disk galaxies.
By tracing particle motion around the spiral arms they  showed that particles move along the arms in the radial direction, migrating towards the outer (inner) radii on the trailing (leading) side of the arm.
On the other hand, \cite{Comparetta2012} found that radial migration may also 
be induced by short-lived transient density peaks (produced by interfering spiral patterns) and it  may be more pervasive than that mediated by the growth and decay of long-lived individual spiral patterns. Migration due to short-lived,
recurent grand design spirals is also found in the simulations of
\cite{2012aAthanassoula}. 
Finally,  the observed diversity of photometric
disk profiles and, in particular, the properties of their outskirts
(see \cite{Bakos2011} and references therein) have been interpreted in terms
of radial migration, either fully \citep{Roskar2008b},  or partially \citep{Sanchez2009}.

In this work, we study the implications of radial migration on the chemical evolution of a barred disk galaxy. We use a N-body+SPH simulation to study the evolution of a disk galaxy embedded in (and interacting with)
a live dark matter halo of 10$^{12}$ \ms, for 10 Gyr; 
an early type disk galaxy with
a strong bar is formed (Sec. 2). We quantify the extent
of radial migration for stars and gas (Sec. \ref{sec:Global}) and we study its
implications for the chemical evolution of the disk
in Sec. \ref{sec:Chemical-evolution}. We find that, despite the important
amount of radial migration occuring in our model, its impact
on the chemical properties is limited. The reason is
the rather flat abundance profile which is established early on in our model, 
due to  the rapid early evolution across the whole disk.

The analysis of the results
allows us to describe the radial displacement of stars in the disk in
a parametrized way. We implement this parametrized description in a ''traditional" detailed
chemical evolution model (including long-lived sources and sinks of elements).
This strategy allows us to overcome the limitations of
the Instantaneous Recycling Approximation (IRA) adopted in the N-body+SPH simulation and to investigate in detail the true impact of radial migration.
We find (sec. \ref{sec:Chemical-post-processing})  that radial migration impacts on chemical
evolution both directly (by affecting the age-metallicity relations, abundance profiles
and metallicity distributions of stars across the disk), and
indirectly, by moving around the long-lived nucleosynthesis sources
and thus altering the abundance profiles of the gas; we show, in particular,
how the radial profiles of O, Fe and D are affected.  Our 
post-processing results show clearly
that the full impact of radial migration on chemical evolution
cannot be evaluated with numerical codes using IRA.

\section{The numerical simulation}
\label{sec:The-numerical-simulation}

The simulation used in this paper was done with the
\textsc{gadget3} code and it
is very similar to simulation 116
described and analysed in \cite{2013Athanassoula}, except that here 
we describe 
the old disc component (\textsc{disk} in the standard \textsc{gadget}
notation) by 200 000 particles and that the softening of the halo and
\textsc{disk} is 100 pc (instead of 50 pc in simulation 116). 

 The adopted simulation 
has four components (\textsc{HALO, DISK, GAS, STARS}), having the
following initial settings: 
\begin{itemize}
\item  A disk made of two initial components that are a gaseous
disk (component GAS), and an old stellar disk (component DISK) having
the same initial azimuthally averaged density distribution. Therefore
the density distribution of the initial total disk is:
\begin{equation}
\rho_{d}(R,z)=\frac{M_{d}}{4\pi h^{2}z_{0}}exp\left(-\frac{R}{h}\right)sech^{2}\left(\frac{z}{z_{0}}\right)
\end{equation} 
Where $M_{d}=5\times10^{10}M_{\odot}$ is the total mass of the disk
(gas and old stars), the initial gas fraction is 0.75, $h=3\, kpc$
is the disk scalelength, and $z_{0}=0.6\, kpc$ is the disk scaleheight.
The DISK component is made of $2.5\times10^{5}$particles having mass
$2.5\times10^{5}\, M_{\odot}$, and has an imposed initial radial
velocity dispersion of $\sigma(R)=100.exp(-R/3h)\, km.s^{-1}$. The
number of particles in this component remains constant during the
simulation. The GAS component features $7.5\times10^{5}$particles
having mass $5\times10^{4}\, M_{\odot}$, which can be partially converted
into new formed stars belonging to the component STARS (this one is
empty at the beginning of the simulation). So the number of particles
in the components GAS and STARS can change.
\item A live spherical halo (component HALO) having the initial density
distribution:
\begin{equation}
\rho_{h}(r)=\frac{M_{h}}{2\pi^{\frac{3}{2}}}\frac{\alpha}{r_{c}}\frac{exp\left(-r^{2}/r_{c}^{2}\right)}{r^{2}+\gamma^{2}}
\end{equation}
where $M_{h}=2.5\times10^{11}M_{\odot}$ is the halo mass, $\gamma=1.5\, kpc$
is the core radius, $r_{c}=30\, kpc$ is the cut-off radius. And $\alpha=\left[1-\sqrt{\pi}exp\left(\gamma^{2}/r_{c}^{2}\right)\left(1-erf(\gamma/r_{c})\right)\right]^{-1}$
is a normalization factor. There are $10^{6}$ particles having mass
$2.5\times10^{5}\, M_{\odot}$in the HALO component.
\item The softening length is 50 pc for all components, and the opening angle
for the tree-code is 0.5.
\end{itemize}
The particles in the component GAS are converted into stars with the prescriptions   (including a threshold on gas volume density) given in
the paper of \cite{Springel2003}, hereafter SH03 (see Sec. 2 and, in particular, eqs. (2) and (23) of that paper), which lead to a satisfactory agreement with the Schmidt-Kennicutt law. Thermal feedback  is also  introduced as in SH03, and the Instantaneous Recycling Approximation (IRA) is used for the chemical enrichment of the gas particles, implemented as described in Sec. 5.3 of SH03. Metal diffusion between neighbouring gas particles is not taken into account. 
No gas infall is considered in this simulation, i.e. the galaxy evolves
as a closed box for 10 Gyr.

\begin{figure}
\begin{centering}
\includegraphics[width=0.49\textwidth]{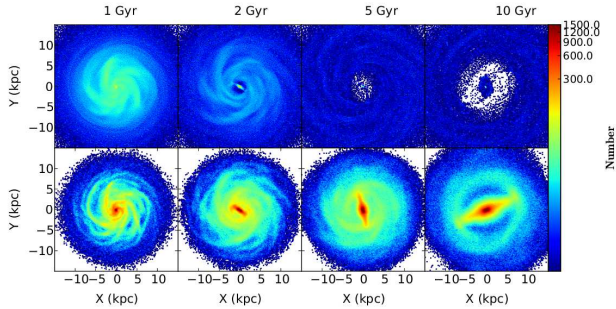} 
\par\end{centering}

\caption{\label{fig:snapshots}Face-on view of the simulated galaxy at 1, 2,
5 and 10 Gyr. Gas-particles are displayed in the upper row and star-particles
in the lower one. The color scale on the right shows the number of
particles in each pixel.}
\end{figure}

Some snapshots of the evolution are shown in Fig. \ref{fig:snapshots}.
Most of the stars are formed during the first $\sim$2 Gyr; in that
period, the gas is depleted and the disk is dominated by the presence
of spiral transient structures. After 2 Gyr, a bar is formed in the
center of the galaxy and grows steadily until the end of the simulation,
when it extends to almost half the size of the stellar disk
The bar
drives gas from the inner Lindblad resonance radius inwards. Evolving locally as in
a closed box (i.e. without being replenished
by infall) the gas is steadily depleted all over the disk, but more
rapidly in the inner regions. Towards the end of the simulation, due
to the combined action of the bar and star formation, the remaining
gas is found mostly in a ring outside the bar which is separated by
the central gaseous concentration by a low density annulus.
The same holds for the stellar disk,
which forms an inner ring (see \cite{Buta1995}  for a definition and description of inner rings).

\begin{figure}
\begin{centering}
\includegraphics[scale=0.4]{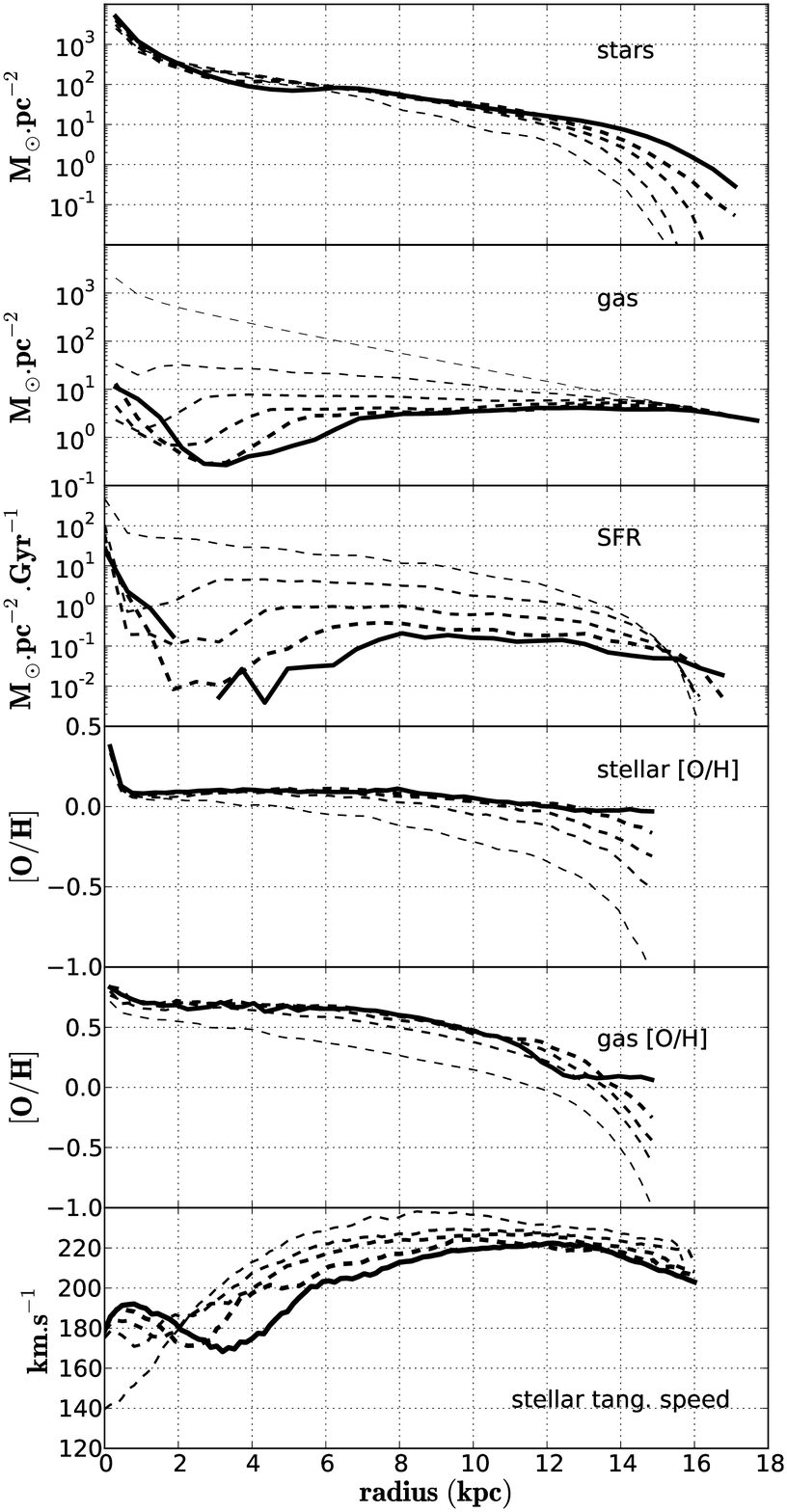} 
\par\end{centering}

\caption{\label{fig:mater-profiles}From top to bottom: azimuthally averaged
radial profiles of the stellar surface density, gas surface density,
star formation rate, average stellar and gas metallicities and rotation
velocity curves. Curves correspond to snapshots taken every 2 Gyr,
with the \textit{thick solid} ones corresponding to the final (10
Gyr) result.}
\end{figure}

The central bar appears to be the most significant
asymmetric structure in the potential, reducing the importance of
spiral arms as it grows (Fig. \ref{fig:snapshots}). Therefore, it
is expected that various dynamical phenomena linked to asymmetries
in the gravitational potential (e.g. disk warming, radial migration,
chaos, resonances, ...), will increasingly be consequences of the
action of the bar, rather than of other structures (such as transient
spiral arms).

Fig. \ref{fig:mater-profiles} displays the evolution of the azimuthally
averaged radial profiles of the surface densities of stars, gas and
star formation rate, as well as of the average stellar and gas metallicities
and the one of stellar tangential velocity. It should be emphasized
that the average surface density profiles provide little information
for the inner disk, which is dominated by the bar. The aforementioned
profiles clearly reflect the inside-out formation of the disk,  with
the gas profiles being much more rapidly depleted in the inner disk than in the outer one, resulting in oxygen profiles becoming flatter with time. The gaseous
profile is mostly depleted in the inner disk, due to both the adopted
SFR law and the action of the bar which gradually produces a spoon-shaped
profile. 
The stars at the edge of the bar can switch from almost circular orbits to elongated ones
(aligned with the bar axis), thus contributing to the bar growth. Once on an elongated orbit,
a star oscillates between central region and the radius of its former circular orbit.
As a result, the stellar profile acquire the same spoon-shape, and it evolves with time as the bar grows longer.  
The SFR profile displays similar features as
the gaseous profile, as expected. The small values of the SFR outside 14 kpc result
in a steep stellar profile beyond that radius. In the absence of infall,
the gas is substantially depleted, only $\sim$8\% of the initial
quantity remaining at the end of the simulation in the whole galaxy and $\sim$13\% in the disk region (outside the bar). One should notice that the final gas metallicity in the outer disk is lower than at previous epochs.
This is a consequence of low-metallicity gas from the region of bar corotation region driven inwards  (see also Sec. 4.1) since  bar corotation is located at R$\sim$15kpc towards the end of the simulation.

\begin{figure}
\begin{centering}
\includegraphics[scale=0.4]{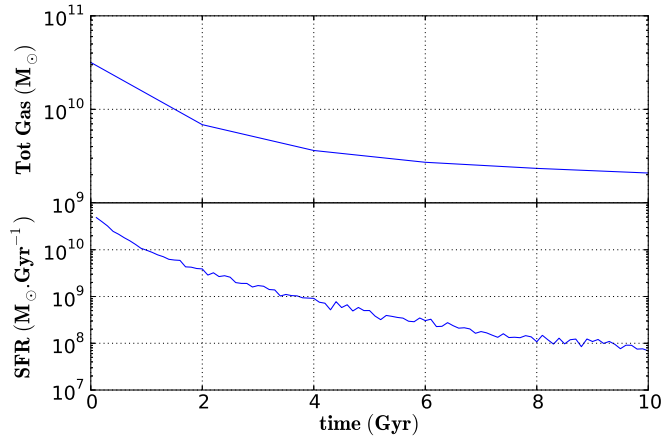} 
\par\end{centering}

\caption{\label{fig:sfr_gas}. Evolution of the total gas amount ({\it top}) and of the total star formation rate 
({\it bottom}) in the simulation.}
\end{figure}

 Fig. \ref{fig:sfr_gas} displays the evolution of the total amount of gas (top) and of the star formation rate (bottom). Due to the adopted criteria for star formation, more than  65\% of the stellar population of the galaxy is  formed
in the first Gyr of the evolution and approximately 85\% is formed {\it in} the first 2 Gyr.

\section{Global behavior of star and gas particles}
\label{sec:Global}

The orbit of a test particle (star) in the potential of a galactic
disk is commonly described, to first order approximation, as the superposition
of a main circular motion (defining the \emph{guiding radius}), and
harmonic oscillations called epicycles. Following Sellwood \& Binney
(2002) we call \emph{blurring} the radial oscillations around the
guiding radius and \emph{churning} the modifications of the guiding
radius.  Churning may occur through resonant interactions of the
star with non-axisymmetric structures of the gravitational potential
(spirals, bar), causing changes in the angular momentum of the stars.
Sellwood and Binney (2002) showed that stars near corotation of a
spiral perturbation may gain (lose) energy and angular momentum as
they fall in the potential well of that perturbation from the leading
(trailing) edge, while conserving the value of their Jacobi constant.
Those changes in angular momentum make them move towards the outer
(inner) disk, where they are deposited
in a new quasi-circular orbit. The process conserves the overall distribution
of angular momentum and does not add random motion, i.e. it does not
heat the disk radially. In contrast,
blurring conserves the angular momentum of individual stars but it
heats radially the disk (the epicyclic radius increases with time).

In a companion paper \citep{Kap2013} 
we analyse in some detail
the behaviour of star particles undergoing churning and, in particular,
the role of bar corotation and its interaction with spiral arms, in that behaviour.
Here we study the impact of both churning and blurring on the chemical
evolution of the disk.
It is instructive to consider first the global behaviour of star and gas
particles in the simulation, by plotting their initial vs. final radius
(for star particles, the initial radius is their birth radius).

\begin{figure}
\begin{centering}
\includegraphics[width=0.45\textwidth]{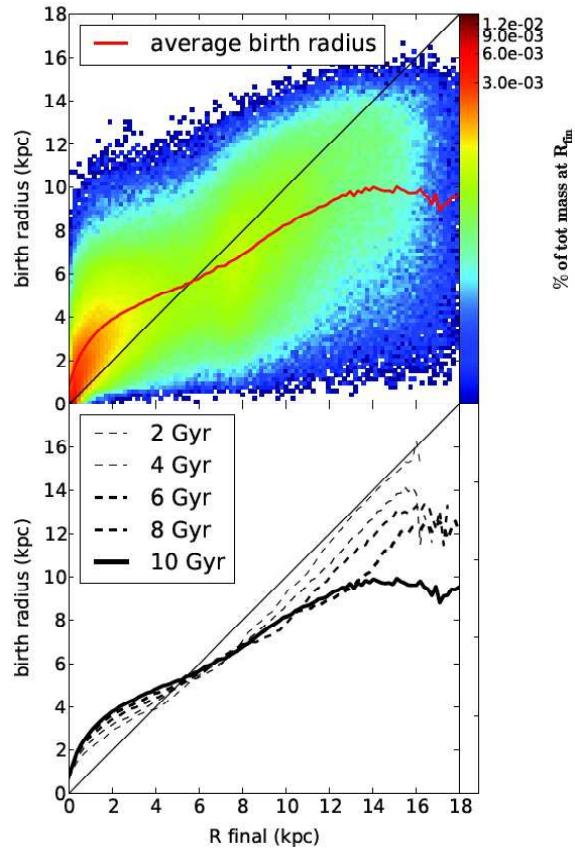} 
\par\end{centering}

\caption{\label{fig:stars-ri-rf} \textit{Upper panel}: birth radius (at any
time) versus final radius (at 10 Gyr) of all star-particles; the color
scale represent the percentage of total stellar mass at final radius,
while the red curve displays the mean birth radius of the stars at
each final radius (in bins 100 pc wide) . \textit{Lower panel}: evolution of the mean birth
radius of the stars at each final radius with snapshots taken every
2 Gyr and with the thickest full curve corresponding to the final
time of 10 Gyr. Non-migrating stars are located on the diagonal black
line (birth radius = final radius). }
\end{figure}

\begin{figure}
\begin{centering}
\includegraphics[width=0.45\textwidth]{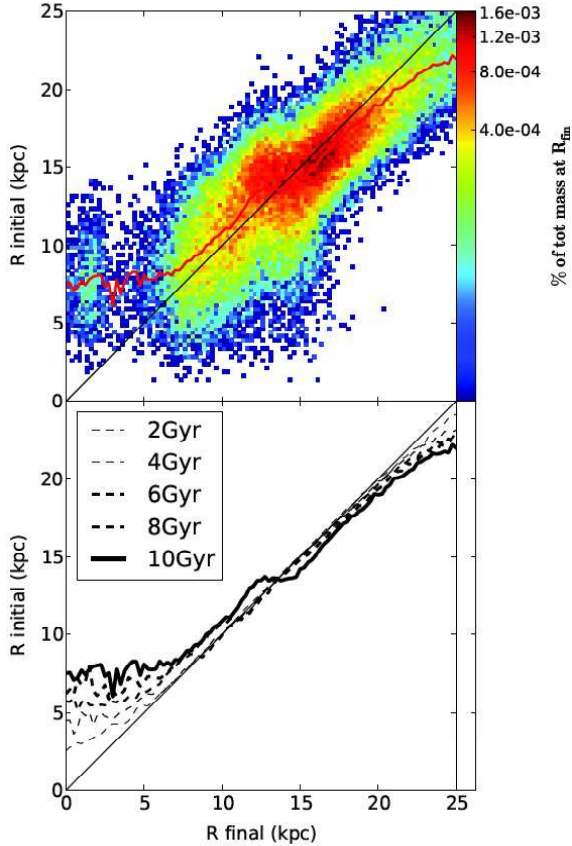} 
\par\end{centering}

\caption{\label{fig:gas-ri-rf} Same as in the previous figure, but for gas
particles. Except that here we consider the initial radius (at t =
0 Gyr), since no new gas particle appear during the simulation time.}
\end{figure}

The results for the star particles appear in Fig. \ref{fig:stars-ri-rf}.
Stars at a given final radius $R_{f}$ originate from a large range
of birth radii $R_{i}$. The distribution of $R_{i}$ vs. $R_{f}$
is not symmetrical with respect to the diagonal implying that it is
not solely due to blurring (=epicyclic motion). The average ratio $R_{f}$/
$R_{i}$ differs significantly across the galaxy: it is smaller than
1 in the inner zones ( $R_{f}<$ 5.5 kpc) and larger than 1 in the
outer ones. Stars in the inner galaxy were born on average a couple
of kpc outwards, whereas stars lying in the outer zone were born several
kpc inwards. The secular evolution of the bar is responsible for the
inwards movements of the stars lying inwards of 5.5 kpc at the end,
because stars on initially near-circular orbits can be captured on elongated orbits
at the edge of the bar, 
implying a mean displacement towards small radii.
The steady increase of the ratio with $R_{f}$ is due to the fact
that there are smaller and smaller amounts of native stars as one
moves to the outer disk: the corresponding average birth radius of
the stellar population at $R_{f}$ is then more and more affected
by the population which migrated from the inner zones.

The results for gas-particles, appear in Fig. \ref{fig:gas-ri-rf}
(upper panel), where the birth radius of the previous figure is replaced
by the initial radius since all the gas particles are present at the
beginning of the simulation. In the upper panel, we distinguish two
regions: inner ($\leq$8 kpc) and outer ($\geq$8 kpc) one.

In the inner region, almost all gas particles come from larger radii
(there is practically no particle which has kept its original radius
since the beginning). As for the stars, the inward movement of the
gas particles lying in the inner zones is due to the bar (\cite{1979ApJ...233...67R,Athanassoula1992dustlanes});
this gas falls to the innermost regions following orbits with an important radial component.
As a result, since the gas forms a continuous fluid (contrary to the
stars), the stream of falling gas creates a shock at each leading
edge of the bar. The gas at radii smaller than the corotation radius, initially
on circular orbits, loses energy and angular momentum each time it
passes through the shock, thus it also falls towards the inner galaxy.
The inflowing gas forms a small central gaseous disk (Fig. \ref{fig:snapshots}
and \ref{fig:mater-profiles}), where the gas density remains high
enough to trigger star formation even at late times (Fig. \ref{fig:mater-profiles}).
The inner galaxy is then constantly depleted of its gas by star formation
and fed by the action of the bar, which brings gas from more and more
large radii as the corotation moves outwards (because of the slowing down
of the bar). As a result, the area between the central gaseous disk
and the corotation of the bar is almost devoid of gas in the end.

In the outer regions (R$>$6 kpc), 
the gas particles are only
slightly affected by the bar corotation radius: particles at radii
smaller than corotation move slightly inwards ($\leq$1 kpc) while
the ones at larger radii move slightly outwards. But the
average initial radius remains close to the diagonal, around which
the distribution of initial radii stays roughly symmetric.

Comparing Fig. \ref{fig:stars-ri-rf} and \ref{fig:gas-ri-rf}, we
see that both stars and gas display strong evolution in the inner regions,
while only the stars display such evolution in the outer galaxy. The
average displacement of gas particles in response 
to gravitational potential asymmetries is
smaller the one of stellar particles, because gas is a continuous dissipative
fluid while stars form a non-dissipative, discrete, one. The bar affects
the gas mainly by driving large amounts of it to the inner regions,
where they fuel star formation. Its role on the behaviour of the stellar fluid is more
diverse, as will be discussed in the next sections.

\section{Chemical evolution}
\label{sec:Chemical-evolution}

One of the major applications of the work of SB02 concerned the age-metallicity relation in the solar
neighborhood. Using a toy model for the evolution of the metallicity
profile of the Milky way disk and a probabilistic description of the
radial migration, the authors showed that a large dispersion can be
obtained in the local age vs. {[}Fe/H{]} relation. They found that
this dispersion can be substantially larger than the corresponding
observational scatter and compatible with the results of the survey
of \citep{Edvardsson+93}. The seminal paper of \cite{Roskar2008a}
revealed other implications of radial migration, concerning the 
metallicity distribution and the stellar metallicity gradients in the disk.  
 
In this section we explore the aforementioned
consequences of radial migration
in our bared disk
model. Our investigation is limited by the use
of just one metal in the numerical simulation,
and by the fact that the Instantaneous Recycling Approximation has
been adopted. For those reasons, standard monitors
of chemical evolution like e.g. abundance ratios (O/Fe etc.) cannot
be used and the results are inaccurate at late times and low gas fractions
(where IRA fails to account for the late return of metal poor gas
from long-lived low-mass stars). Still, it is instructive to explore
some of the implications of that approximate treatment in order to
get some insight into the effects of radial migration on chemical
evolution.

\subsection{The age-metallicity relation}

We assume in the following that the unique metal
in the simulation represents oxygen, because it constitutes almost
half of the solar metallicity and because its evolution is 
described by IRA in a satisfactory manner in many cases, being a product of short-lived massive
stars. We are fully aware that at late times IRA may over-predict
oxygen abundances, but we are interested mostly in relative, not absolute,
values of the abundance.

\begin{figure}
\begin{centering}
\includegraphics[scale=0.4]{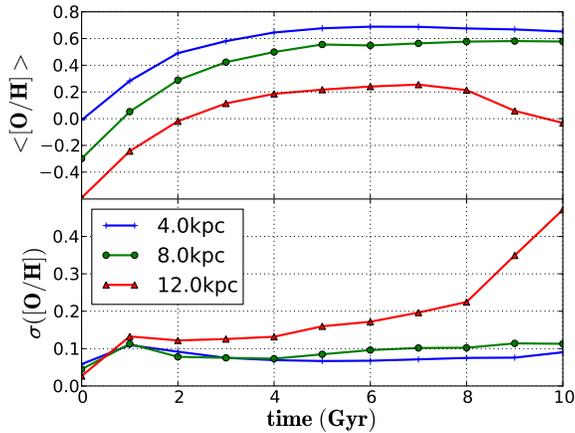} 
\par\end{centering}

\caption{\label{fig:gas-metallicity}Evolution of the azimuthally averaged
gas metallicity {{[}}O/H{{]}} (\textit{upper panel}) and of
the corresponding dispersion (in dex, \textit{lower panel}) at three
different galactocentric zones of width $\Delta R$=1 kpc: 4 
(solid blue), 8 (green circles) and 12 (red triangles) kpc. }
\end{figure}

Fig. \ref{fig:gas-metallicity} (upper panel) displays the evolution of the azimuthally
averaged metallicity in three different galactocentric radii, located
at 4, 8 and 12 kpc, respectively. Metallicity increases rapidly in
the first couple of Gyr and more slowly (less than a factor of 2 or 0.3 dex)
in the remaining evolution. This applies to all radii, albeit
with a small delay in the early stages, due to the inside-out formation
of the disk (see 5th panel from the top in Fig. \ref{fig:mater-profiles}).
The most conspicuous feature in Fig. \ref{fig:gas-metallicity} is
the decline of the metallicity at 12 kpc in the last 2 Gyr of the
evolution which is, in principle, unexpected. The origin of that feature
is understood after inspection of Fig. \ref{fig:mater-profiles}:
in the last 2 Gyr of the evolution, bar corotation is located at $\sim$12
kpc and drives inwards gas from the outer regions; that gas is metal
poor because the metallicity profile declines rapidly outside 12 kpc
(due to weak stellar activity at R$>$12 kpc),
and dilutes the oxygen abundance at that radius while at the same
time flattens the abundance profile in the region between 12 and 16
kpc.

The lower panel of Fig. \ref{fig:gas-metallicity} shows the dispersion
in the gas metallicity in those same zones. Dispersion is fairly small
(less than $\sim$0.1 dex) during the whole evolution in the 4 kpc
and 8 kpc zones. In contrast, it becomes important in the 12 kpc zone
in the last couple of Gyr, for the same reason as the decline in metallicity
of that zone, namely the dilution with metal poor gas from the outer
disk. This gas is driven inwards by the bar corotation and does not
exchange  metals with local gas, because of the simplified treatment
of metals in 
the adopted version of the {\t Gadget} code. Thus, although the effect
of the declining metallicity discussed in the previous paragraph is
real (at least in the framework of our model), the effect of a late
increasing dispersion is a numerical artifact, which impacts also
on the results of the stellar metallicity dispersion as we shall see
in the next paragraph%
\footnote{We are currently implementing a considerably improved module of chemical
evolution in the GADGET code, which contains elements from short-lived
and long-lived sources - including Fe from SNIa -, and accounts for
mixing of gas phases.%
}.

\begin{figure}
\begin{centering}
\includegraphics[width=0.5\textwidth]{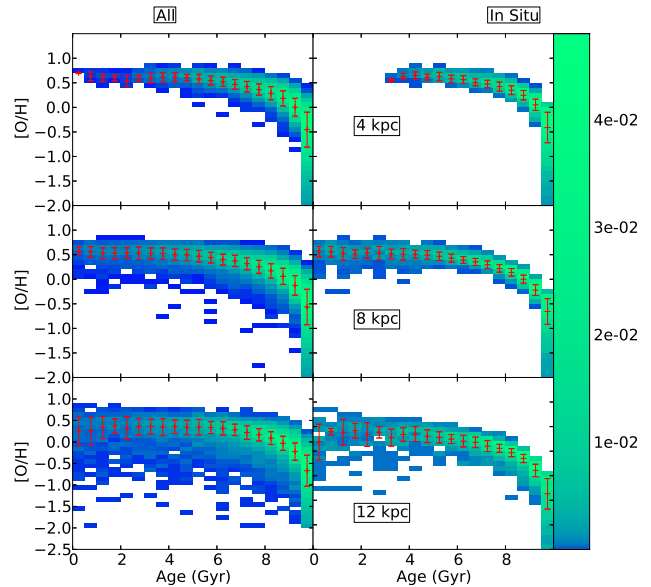} 
\par\end{centering}

\caption{\label{fig:stars-age-met}Stellar age-metallicity relation for three
zones at galactocentric radii R=4, 8 and 12 kpc, respectively (from
top to bottom) and width $\Delta R=\pm1kpc$. Results are displayed
for all stars present in the zone at 10 Gyr (left panels) and for
stars which were formed in that zone and are still there at 10 Gyr
(\textit{in-situ} stars, right panels). In all panels, at each age
bin ($\Delta$(Age)=1 Gyr) the metallicity distribution is fitted
by a Gaussian function, and its peak and standard deviation are displayed
with (red) dots and error bars, respectively.}
\end{figure}

The stellar age-metallicity relation at 4, 8 and 12 kpc is displayed
in Fig. \ref{fig:stars-age-met}, for all the stars found at 10 Gyr
in those zones (left panels) and for those stars only that have been
formed in those same zones (right panels). In all zones, metallicity
remains flat for stars younger than 6 Gyr, as expected, because of
insignificant star formation during that period. For stars born \textit{in-situ},
dispersion is small and follows the corresponding gaseous dispersion
(see Fig. \ref{fig:gas-metallicity}), except in the first couple
of Gyr where the gas metallicity increases rapidly: since our stellar
age bins have a width of 1. Gyr, we obtain a large range of metallicity
values in the first bins (and consequently a large dispersion) even
though the corresponding "instantaneous" 
dispersion in the gas is small.  It should be noticed here that the term ''dispersion" has not exactly the same meaning in the case of gas and stars: for the gas, it means {\it instantaneous dispersion} (at any given time), but for the stars
it always concerns a given {\it age-interval} (here taken to be of 1 Gyr); as a result, independently of any radial migration, the latter is always larger than the former: stars of age 4($\pm$0.5) Gyr have a larger metallicity dispersion than the
gas had 4 Gyr ago.

At late times, metallicity evolves little and \textit{in-situ} formed stars present
smaller dispersion. Radial mixing, either through churning or blurring,
increases that dispersion, albeit by modest amounts. We confirm the
trend originally found in SB02 of smaller dispersion with decreasing
age: it is due to the fact that younger stars have less time to migrate
from far away regions. However, the overall effect is smaller in our
case because the metallicity gradients we obtain are substantially
flatter than in SB02 at all ages.  At very late times, dispersion increases in the outer regions (see panels for 8 and 12 kc), for a reason independent of radial migration: The bar corotation progressively moves outwards, driving inwards metal poor gas from the outer disk which is mixed (but not completely) with metal richer gas in the regions from 8 to 12 kpc. This is seen in Fig. \ref{fig:mater-profiles} (the gas metallicity profile of the outer disk flattens at 10 Gyr from metal-poor gas driven inwards) and in Fig. \ref{fig:azimuth}  (bottom left panel, with important azimuthal variations of gas metallicity at 12 kpc). It is these variations in gas metallicity( from incomplete gas mixing from outer regions) that drive the large dispersion in stellar metallicity in the outer regions at late times.

\begin{figure}
\begin{centering}
\includegraphics[width=0.5\textwidth]{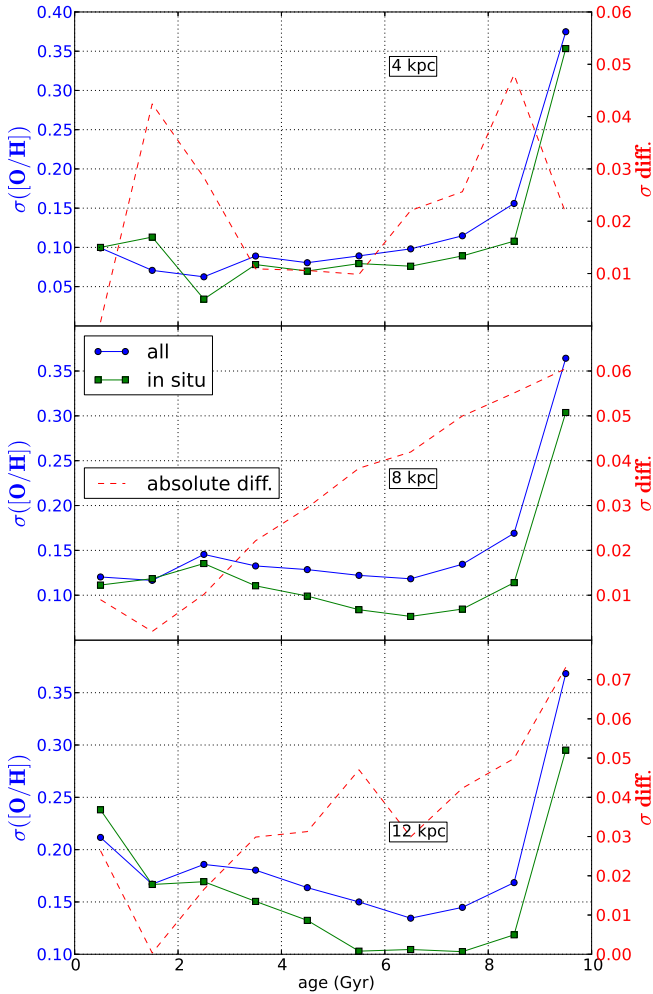} 
\par\end{centering}

\caption{\label{fig:sigma-age-met} Evolution of the stellar metallicity dispersion
in three zones of galactocentric radius $R$=4, 8 and 12 kpc and width
$\Delta R=\pm$1 kpc. The dispersion is calculated as one standard
deviation $\sigma([O/H])$ around the mean value. The blue curves
(circles) correspond to all stars present in the zone and the green
curves (squares) to stars formed within the zones; both curves read
on scales on the left. The red (dashed) curve represents the difference
of the two others $\sigma_{dif}=\left|\sigma_{all}-\sigma_{insitu}\right|$ and
is read on the right axis.}
\end{figure}

Those results are quantitatively illustrated in
Fig. \ref{fig:sigma-age-met}, which shows that dispersions generally
decrease with time  and that
the fractional dispersion difference between all the stars and the
\textit{in-situ} formed ones is also a decreasing function of time  (except in the case of stars at 8 and 12 kpc, due to the aforementioned reasons).

\subsection{Metallicity distributions}

\begin{figure}
\begin{centering}
\includegraphics[width=0.5\textwidth]{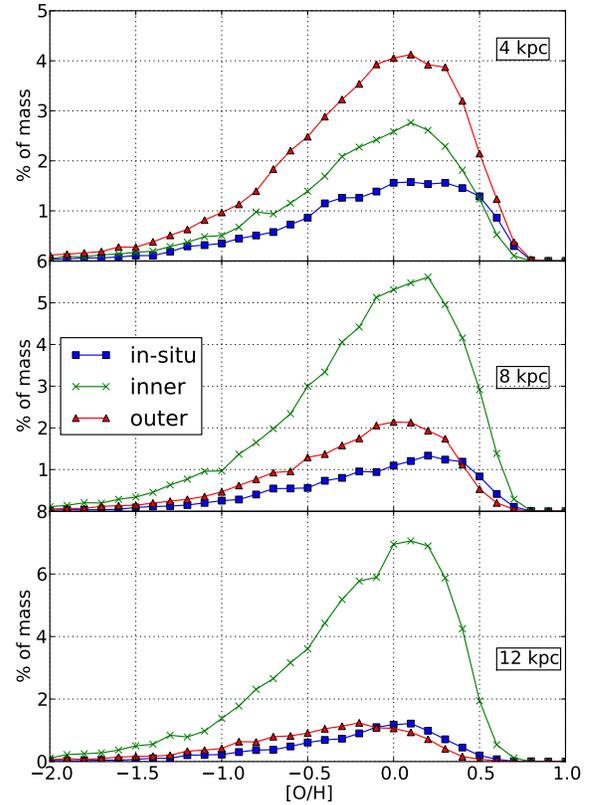} 
\par\end{centering}

\caption{\label{fig:met_distrib_io} Contributions to the metallicity distributions
of three radial zones (R=4, 8 and 12 kpc, from top to bottom, and
width $\Delta R=\pm$1 kpc) from stars born \textit{in situ} (squares
and blue curves), coming from the inner galaxy (crosses and green
curves) and from the outer galaxy (triangles and red curves). The
width of the metallicity bins is 0.1 dex. }
\end{figure}

As discussed in \cite{Roskar2008a}, radial mixing reshuffles
the metallicity distributions of stars across the disk. Our simulation
shows clearly this effect, as can be seen in Fig. \ref{fig:met_distrib_io}.
In each one of the $\Delta R$=1 kpc wide zones centered on 4, 8 and
12 kpc, respectively, stars born \textit{in situ} constitute a minority
at all metallicities. The stellar population at 4 kpc is dominated
by stars from outer zones while those at 8 and 12 kpc are dominated
by stars from the inner disk.

 Due to the rapid metallicity evolution,
stars originating in different
regions and ending in the same zone cover the whole metallicity range
but {\it differ little in their average metallicity}. The small differences in the peak metallicities are explained by the fact that all zones in our model evolve practically as closed boxes, since gas particles do not suffer significant radial displacement and there is no gaseous infall. In that case the peak of the metallicity
distribution corresponds to the stellar yield, e.g. \citep{Prantzos2012}, 
which is fixed in the model. The largest differences occur in the zones at 8 and 12 kpc, where stars originating in the
inner and the outer zones differ, on average, by 0.1 dex in metallicity.
 Combined to the fact that most of those stars are formed quite early 
(first 2 Gyr) and have enough time to migrate all over the disk, this produces 
a negligible change in the peak of the metallicity distribution.  

\begin{figure}
\begin{centering}
\includegraphics[width=0.5\textwidth]{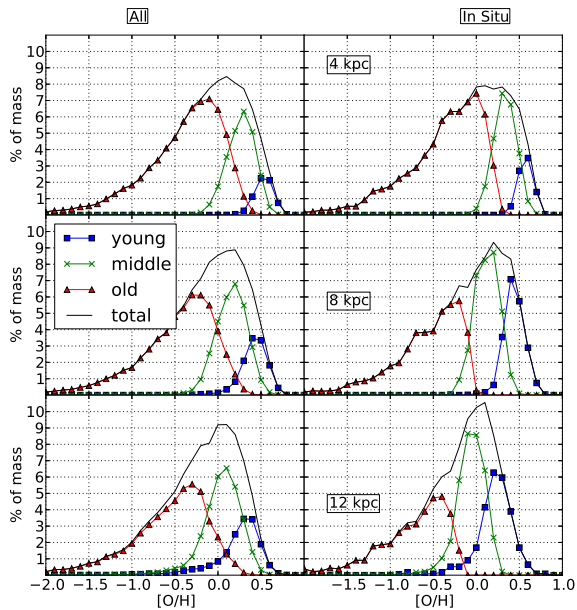} 
\par\end{centering}

\caption{\label{fig:met_dist_all}Metallicity distributions in three regions
(of radius $R$=4, 8 and 12 kpc, from top to bottom), for all the
stars found in those zones (left) and for the stars born \textit{in
situ} (right). In each panel, three classes of stellar ages are displayed:
''young" ($<$7 Gyr, squares and blue curves), ''middle-aged" 
(7-9 Gyr, crosses and
green curves) and ''old" ($>$9 Gyr, triangles and red curves); their sum (total
metallicity distribution) is indicated by the black curves. The width
of the metallicity bins is 0.1 dex. }
\end{figure}

 For that same reason, the width of the metallicity distributions
is barely modified between stars born \textit{in situ} and all the
stars found in a given zone, as can be seen by comparing the black
solid curves between left and right panels in Fig. \ref{fig:met_dist_all}.
On the other hand, radial mixing does change the ratio between young
and old stars in a given region, as can be seen in that same figure:
as one moves outwards, from 4 to 12 kpc, the fraction of ''young" stars
(here defined as those younger than 7 Gyr) born \textit{in situ} becomes
more and more important with respect to the one of ''old stars" (older
than 9 Gyr), because of the inside-out star formation; however, if
all stars are considered, then ''young" stars are always a minority,
even at 12 kpc (right panels in Fig. \ref{fig:met_dist_all}), because most of star 
formation occurs in the first couple of Gyr in our model.

\subsection{Evolution of abundance profiles}

The shape of the abundance profiles of disk galaxies constitutes a
key diagnostic tool of their evolution. It has been realized long ago,
with simple (independent-ring) models,  that
the inside-out formation of disks produces generically profiles 
in the gas and young stars which flatten with time 
\citep{1989MattFranc,1995PrantzosAubert,1999BoissierP,2000HouBP}. 
This was
confirmed by chemo-dynamical simulations, e.g. \citet{2003SamlandGerhard}. 
In a recent study comparing
various codes of disk evolution for Milky Way type disks, 
\citet{2012Pilkington}) find that: i) the gradient of oxygen 
may vary widely from one simulation to another, but in most cases it is
substantially larger than observed in the Galaxy and ii) in most
cases, the oxygen profile flattens with time.

\begin{figure}
\begin{centering}
\includegraphics[width=0.5\textwidth]{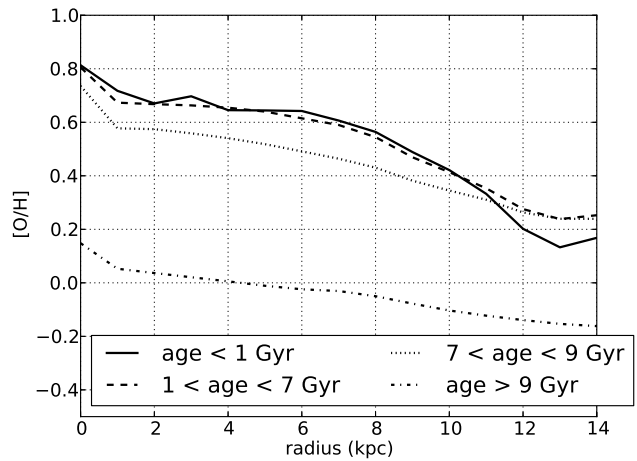} 
\par\end{centering}

\caption{\label{fig:metal_pop} Abundance profiles of 
stars in various age bins. After radial mixing, profiles appear to be flatter
for old stars, in contrast to what happens with the gaseous
profiles.}
\label{final-met-pops}
\end{figure}

On the basis of such considerations, it was expected that observations of stars
of various ages across the disk of the MW could reveal the shape of the abundance profile as a function of time 
allowing one to probe the evolution of that profile
and thus to draw conclusions about the local history of star formation rate vs. infall and/or radial inflows of gas. 
Although most studies of disk chemical evolution agree that profiles flatten with time,
some models (e.g. \citet{CMR01}) based on a different assumptions conclude that profiles steepen with time.

Such observations have been conducted over the years,  using
mostly planetary nebulae to trace the past evolution of abundance 
profiles.
In particular, \citet{MC09} find that the abundance profile of oxygen flattens with time, thus supporting  quantitatively the findings of 
\citet{2000HouBP}. However, \citet{2010StangelH} find the opposite trend, namely a steepening of oxygen profiles with time.
It appears that the systematic uncertainties in ages and distances of those sources are so large at present that they do not allow for a robust evaluation of past abundance gradients. The situation may change in the future, with improvement in distance estimates and the use of proxies for the age, e.g. the [$\alpha$/Fe] ratio
\citep{Cheng2012a}.

\cite{Roskar2008a} realized that one of the consequences of radial migration is to reduce and even  inverse the abundance gradient of a stellar population across the galactic disk: an early steep gradient may be subsequently
erased by migration of  old stars towards the outer regions.
Thus, even in the absence of any systematic uncertainties,  abundance determinations in planetary nebulae
would be of little help in  revealing the past abundance profile of the disk.

We present in Fig. \ref{final-met-pops} our profiles in stars of various age bins
It can be seen that, contrary to Fig. 2, where the gas abundance profile systematically flattens with time, the stellar average abundance 
profile appears to be flatter
for older stars. This confirms the finding of \citet{Roskar2008a} and implies that even accurate observations of present days 
abundance profiles of stars of various ages cannot reveal the past
history of the true abundance profile of the disk.

We notice, however, that the implications of this finding are not
totally negative because, 
{\it in the case of a monotonic evolution of the profile,}
the true abundance gradient of a stellar population at the time of its formation {\it should have been steeper  than observed today }
(since radial migration always flattens it). 
If  the presently observed abundance profile of old stars
is steeper than the one of the  gas (as claimed
 by e.g. \cite{MC09}) then
it can be safely inferred that the  disk evolved from a steeper to a flatter profile. This is by itself an important conclusion, albeit 
in a qualitative level.

\begin{figure}
\begin{centering}
\includegraphics[width=0.5\textwidth]{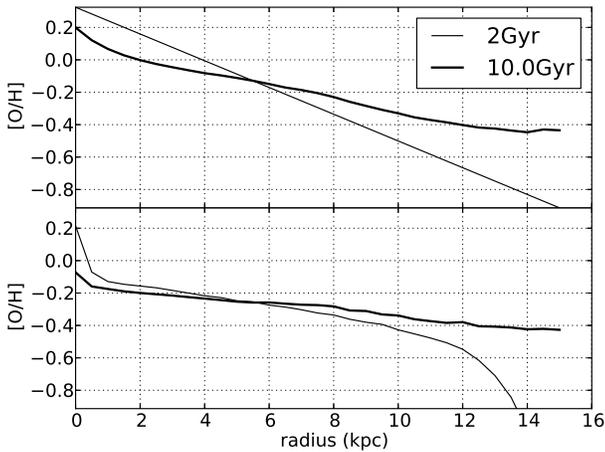} 
\par\end{centering}

\caption{\label{fig:imp_gra} 
Evolution of the stellar metallicity gradient in the case of our
simulation (bottom) and of the same simulation with an imposed initial 
gradient of -0.08 dex/kpc (top). The stellar population after time t=2 Gyr is
displayed with thin curves in both panels; 
that same population, at the end of the simulation (i.e. aged of 8 Gyr)
appears in thick curves.}
\end{figure}

The strong early star formation of our disk galaxy leads to small 
early abundance gradients and thus
minimizes the effects of radial migration, despite the fact that the stars
have time enough (8 Gyr, on average) to migrate.
Thus, the average metallicity profile of stars after the first 2 Gyr (which
constitute about 90\%  of all stars) in our simulation has a slope of
dlogZ/dR=-0.03 dex/kpc. At the end of the simulation (10 Gyr) 
those same stars display a slope of -0.02 dex/kpc, i.e. radial migration
has modified the slope of the abundance profile by only -0.01 dex/kpc.

In order to evaluate  the effect of a steep initial metallicity profile
(while keeping the same dynamical evolution)
we imposed an artificial gradient of -0.08 dex/kpc to all stars formed in the
first 2 Gyr and we assumed that the disk evolved
exactly as in our simulation.
The results appear in Fig. \ref{fig:imp_gra} (top panel) 
and are compared to those
of the original  simulation (bottom panel).
It can be seen that the slope of the final abundance profile is of -0.04 dex/kpc,
i.e. the same amount of radial migration modified the slope by -0.04 dex/kpc,
instead of -0.01 dex/kpc in the original simulation.
We conclude that the effect of radial migration on the final abundance
profile of stars depends not only on the strength of the various churning
mechanisms, but also on the overall history of the disk (star formation vs. infall and resulting chemical evolution). 

\subsection{Azimuthal variations of metallicity}
\label{Subsec:Azimuth}

The question of azimuthal variations in either the gaseous or stellar 
metallicity of a disk galaxy has been addressed by various authors 
either from the observational (\cite{Li2013} and references therein)
or from the theoretical point of view, e.g. \cite{2013DiMatteo} and references therein.
In particular, \cite{2013DiMatteo} explored the possibility of 
azimuthal variations
in old star composition as signatures of radial migration, in the case of a barred
galaxy. In their controlled experiment (N-body with the SPH part switched off, i.e. 
no star formation) they imposed initial metallicity profiles and studied the
azimuthal distribution of star particles after 4 Gyr. They found azimuthal
metallicity variations depending on the initial metallicity profile and 
persisting during the whole period of bar activity.

\begin{figure}
\begin{centering}
\includegraphics[width=0.5\textwidth]{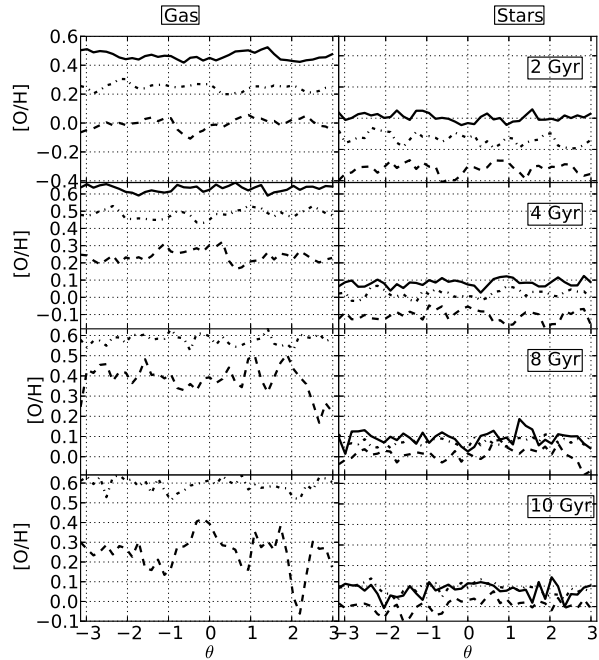} 
\par\end{centering}

\caption{\label{fig:azimuth} 
Azimuthal variations of metallicity in four different times (2, 4, 8 and 10 Gyr, from top to bottom)
and in three different galactocentric distances: 4, 8 and 12 kpc 
(solid, dotted, dashed,  respectively), for the gas (left) and for the
stars (right).}
\end{figure}

Our results concerning the azimuthal metallicity of  gas and stars are displayed in
Fig. \ref{fig:azimuth}. It can be seen that in early times
azimuthal variations are small in both gas and stars (0.1 dex at maximum).
At late times, as the bar corotation moves outwards, metal poor gas is driven
inwards from the outer disk, creating local 
azimuthal variations of up to 0.3 dex
(a factor of 2) in the gas metallicity at 12 kpc. 
As already discussed for Fig. 2, this important gas inflow reduces the average radial metallicity of gas in the outer disk.
However, the
bulk of the stellar population in all radii is formed in early times
and despite substantial radial migration it shows no significant azimuthal
variations in its metallicity, because the metallicity gradient is always
small. If a steep metallicity profile is imposed (as in Fig. \ref{fig:imp_gra})
stronger azimuthal variations in the stellar population are obtained, but they never exceed the corresponding variations in the gas metallicity.

We notice that important azimuthal variations in the gas metallicity can
also be obtained in the case of local infall of metal poor gas (infall is not
included in our simulation). In an analoguous way, a merger with a metal poor
satellite could also create azimuthal variations in the stellar metallicity.
For those reasons, it appears that azimuthal variations of metallicity in either
gas or stars do not provide unambiguous information about radial migration.

\section{Chemical post-processing and the implications of non-IRA}
\label{sec:Chemical-post-processing}

As discussed in Sec. 4, the analysis of the chemical evolution part
of our simulation is limited by the use of IRA and of a single metal.
Here we show that it is possible to overcome this limit and gain considerable
more insight, by post-processing the
evolution of the simulated disk with a simple, classical model of
galactic chemical evolution including a much more elaborated chemistry.
In this section we first establish a parametrized description of the
churning+blurring processes in the disk of our simulation. We then
implement it in a classical chemical
evolution model for that same galaxy, first with IRA, in order to
check whether the results of the numerical simulation - which uses
IRA - are satisfactorily reproduced. We show that this is indeed the
case. We drop then the IRA and run the same model by introducing more
chemical elements and, in particular, Fe from SNIa. This procedure
allows one to exploit in detail the chemical evolution of the system
(by introducing more metal sources or different prescriptions for
the rates for e.g. SNIa), once the successful description of churning+blurring
through a parameterized scheme is established.

\subsection{Parametrization of churning+blurring}

Up to now, two different types of parameterization of radial mixing
have been introduced in the literature. SB02 adopted a global
mixing scheme, in which one assumes that a star born at radius $R_{0}$
at time $t$ may be found at time $t_{0}$ (i.e. after time $\tau=t_{0}-t$)
in radius $R_{f}$ with a probability $P(R_{0},R_{i},\tau)$ given
by a Gaussian function 
\begin{equation}
P(R_{f},R_{O},\tau)\ =\ (2\pi\sigma_{\tau}^{2})^{-1/2}\exp(-\frac{(R_{f}-R_{0})^{2}}{2\sigma_{\tau}^{2}})\label{eq:prob}
\end{equation}
 where $\sigma_{\tau}$ is the 1 $\sigma$ dispersion in the radial
displacement of the particles from their birth place $R_{0}$ after
time $\tau$. SB02 adopt, for illustration purposes an expression
for $\sigma_{\tau}$ which includes two terms accounting for churning
and blurring, respectively; notice that their term for churning is
not symmetric with respect to particle exchange (i.e. it depends explicitly
on $R_{0}$) but this does not have to be generically the case.

On the other hand, \cite{SB09} treat separately blurring
and churning. For the former, they make some assumptions about the
radial dependence of the radial velocity dispersion of stars. For
the latter, they adopt a local scheme,
in which only stars from second-nearest neighboring zones can exchange
places during a time-step, with a probability adjusted to reproduce
some observables in the solar neighborhood, such as the metallicity
distribution. According to their scheme, not only stars but also cold
gas (molecular) is affected by churning. However, it seems improbable
that gas - which is dissipative - behaves as the collisionless fluid
of stars. Furthermore, molecular gas is bound in molecular clouds
with lifetimes of only 10$^{7}$ yr, i.e. too short for any appreciable
radial displacement. The results of our numerical simulations support
these considerations, as discussed in Sec. 3 (see also Fig. \ref{fig:gas-ri-rf}),
so we shall ignore here any radial migration of gas particles; we
shall see that this approximation is valid for most of the disk, but
not inside the bar, which drives rapidly gas towards the central regions.

In this work, we adopt an approach similar to the one of SB02, since
their parametrization is supported by the analysis of numerical simulations
by Brunetti et al. (2011) who analyzed a star-only controlled simulation
performed with the Gadget-2 code. Although they had no gas or star
formation in their simulation, their conditions are not very different
from ours, since in our case star formation occurs essentially in
the first couple of Gyr. Brunetti et al. (2011) found that the radial
displacement of stars in their simulation can be described by Gaussian
functions simulating a diffusion process. They caution that modeling
the stellar migration as a diffusion process is valid only for time
intervals less than the diffusion time-scale, which they estimate
from the simulation results to be of the same order as the rotation
period. In other terms, radial migration can be described as a diffusion
process with diffusion coefficients depending both on time and (original)
position.

\begin{figure}
\begin{centering}
\includegraphics[angle=-90,width=0.5\textwidth]{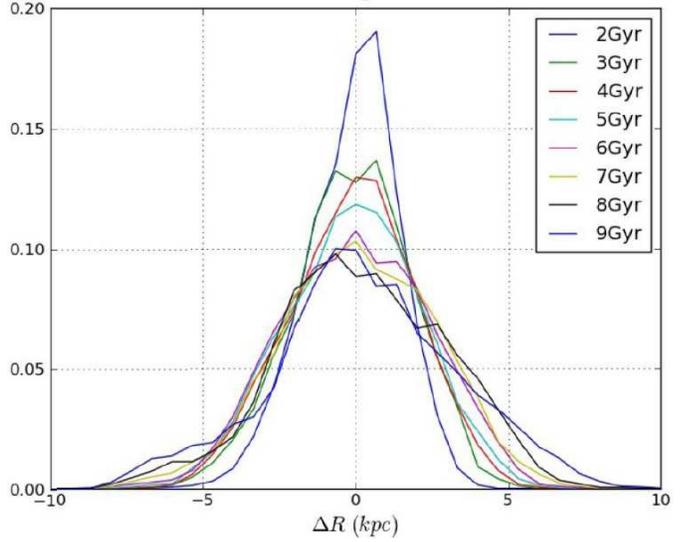} 
\par\end{centering}

\caption{\label{fig:RadDiff} Radial positions of particles born at radius
$R_{0}$=8kpc between 0 and 1 Gyr after times $\tau$ spanning the range of 2 to 9 Gyr. }
\end{figure}

We first follow the positions of all the stars of the numerical simulation
born at a given radius $R_{0}$ as a function of time. As displayed
in Fig. \ref{fig:RadDiff} those stars are found at later times $t$
at various positions $R_{t}$. The distributions of particles as a
function of the position can be approximated by Gaussians with widths
generally increasing as a function of time. Similar behavior characterizes
the radial positions of stars born in that same radius $R_{0}$ at
later times or at different original radii. By constructing a sufficiently
dense grid in $R_{0}$ and $t$, we find then that the evolution of
the radial dispersion of stars can be described by expressions similar
to the one of Eq. \ref{eq:prob}, where the widths are given by:

\begin{equation}
\sigma_{\tau}(R_{0})\ =a(R_{0})\tau^{N}+b(R_{0})\label{eq:FWHM}
\end{equation}
Our fitting procedure produces values of N in a narrow range N=0.4-0.5 at all radii
and we fixed here N=0.5.
 Using a least square method, we get 
 the values of $a(R_{0})$ and $b(R_{0})$ at each radius $R_{0}$: 
\begin{equation}
a(R_{0})\ =-6.67e^{-2}R_{0} + 2.75\label{eq:fit-a}
\end{equation}
\begin{equation}
b(R_{0})\ =-2.26e^{-1}R_{0} + 2.71\label{eq:fit-b}
\end{equation}

We implement this description of radial migration in a 1D code of
chemical evolution with independently evolving annuli \citep{1999BoissierP}.
The initial configuration contains a dark matter halo of 10$^{12}$
M$_{\odot}$ with a NFW profile and a gaseous disk of 5 10$^{10}$
M$_{\odot}$ with an exponential scalelength of 2.7 kpc, i.e. the
same initial conditions as the N-body+SPH simulation. The disk annuli
evolve as closed boxes, assuming azimuthal symmetry and IRA. The local
star formation rate is assumed to be $\Psi(R,t)=\nu\frac{V}{R}\Sigma_{Gas}^{1.5}$,
with $\nu$ adjusted as to reproduce the final gas profile of the
simulation. During a time-step $dt$ a mass of stars $m_{S}(R_{0},t)$=$\Psi(R_{0},t)dt$
is created at radius $R_{0}$. In subsequent time-steps, that mass
undergoes radial migration to other zones $R$ according to the adopted
probabilistic description of Eqs. \ref{eq:prob} and \ref{eq:FWHM}.
Obviously, if the final profiles of gas, gas metallicity and SFR of
the 1D simulation match the corresponding final profiles of the N-body+SPH
simulation, the chemical evolution part of the former simulation can
be considered as a succesfull description of the latter. And if the
final profiles of stars and stellar metallicity \textit{as well as}
the initial vs. final radii of the two simulations match each other,
then the adopted probabilistic description scheme can be considered
as a successful description of the radial migration obtained by the
N-body code.

\begin{figure}
\begin{centering}
\includegraphics[width=0.5\textwidth]{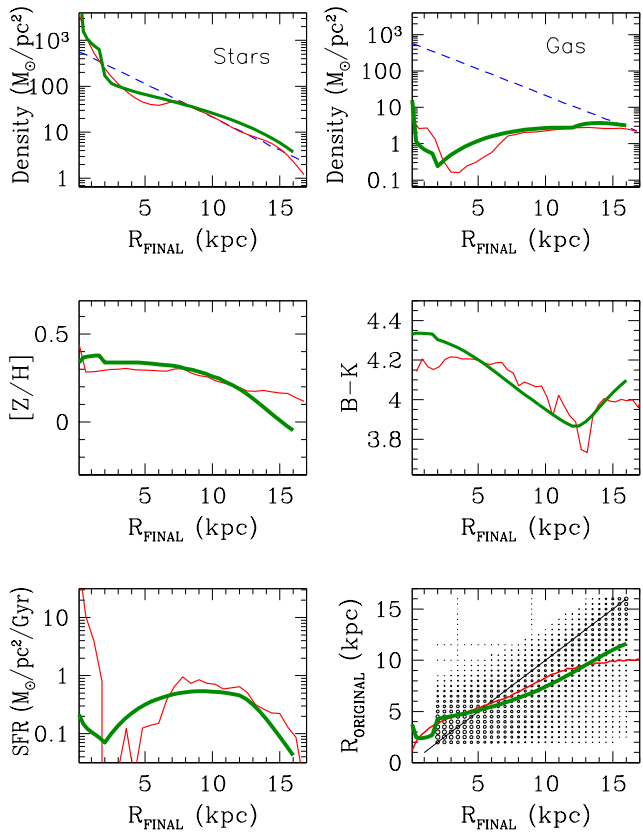} 
\par\end{centering}

\caption{\label{fig:NumvsAnal} Results of the 1D chemical evolution model
with IRA and parametrized description of the radial migration, as
developed in Sec. 5.1, compared to those of the N-body+SPH calculation.
In all panels, {\it thin curves} correspond to final profiles (at 10 Gyr)
obtained with GADGET and {\it thick curves} to corresponding profiles obtained
with the parametrized description and the semi-analytical model. \textit{Top left:} Stellar surface density and
\textit{top right:} Gas surface density ; in both panels the {\it dashed curve}
corresponds to the initial gaseous profile of the disk.
\textit{Middle left:} metallicity profile; \textit{middle right:} B-K colour profile.
\textit{Bottom left:} Star formation rate profile. \textit{Bottom right:}
Birth radius vs. final radius for all star particles (to compare with Fig. 3).
 }
\end{figure}

The results of our 1D calculation appear in Fig. \ref{fig:NumvsAnal},
where they are compared with those of the N-body+SPH simulation. It
can be seen that, the final profiles of gas and star formation are
reasonably well reproduced by the parametrized simulation, albeit
with a sizable difference in the bar region (from 2 to 6 kpc). The
reason for that discrepancy is that we do not consider any radial
motion of the gas, an approximation which accounts well for the relation
of $R_{0}$ vs $R_{f}$ of gas particles in most of the disk (see
Fig. \ref{fig:gas-ri-rf}), but not in the region of the bar, which
drives gas inwards. This also accounts for the small discrepancy of
the final SFR observed in that same region. The metallicity of the
gas is well reproduced over the whole disk.

The aforementioned features are obtained in the independent ring approximation
adopted for the gas. The results for the final profiles of stars,
stellar metallicity and colour depend also on the adopted prescription
for radial migration. In Fig. \ref{fig:NumvsAnal} it is seen that
over a 12 kpc region (from 3 to 15 kpc) the curves of the average initial
$R_{0}$ vs final $R_{f}$ radius of the stars between the two simulations
differ by an amount smaller than the difference between $R_{0}$ and
$R_{f}$ in the N-body+SPH simulation (a difference which reaches
almost 5 kpc in the outer disk). This suggests that the adopted description
of radial migration manages to reproduce reasonably well the effect.
Furthermore, the final profiles of average stellar metallicity and
B-K colour are also well reproduced; in particular, we obtain the
upturn of B-K around 13 kpc, as in the N-body+SPH simulation. Taking
into account the extreme simplicity of our formula (using Gaussian
functions with a regular time dependence, while the true situation
is more complicated) we consider the overall result as fairly successful.

In summary, we have shown that a simple model of galactic chemical
evolution, augmented with a simplified description of radial migration
\textit{a la} SB02, can reproduce fairly satisfactorily the results
of a full N-body+SPH calculation, once the diffusion
coefficients have been determined from the latter simulation.
This opens the way for realistic post-processing of N-body simulations,
with simple chemical evolution models including many more chemical
elements and nucleosynthetic sources (that were neglected  in the N-body simulation). 
For instance, one may consider
other elements than the single case of oxygen considered here, like
e.g. Fe from both core collapse and thermonuclear supernovae; it is
also possible to study the evolution of the system dropping IRA and
considering the finite lifetimes of stars, allowing one to consider
the evolution of s-elements or deuterium. We illustrate some of those
possibilities in the next section.

\subsection{Implications for chemical evolution}

We run the same model as in the previous section, considering the
finite lifetimes of stars as well as Fe from SNIa. We adopt the stellar
lifetimes of \cite{1992Schaller}, the stellar yields \cite{1995WW}
and the prescription of \cite{2005Greggio} for the rate
of SNIa, which accounts for single degenerate dwarves as progenitors
of those objects; each SNIa is assumed to eject 0.7 M$_{\odot}$ of
Fe. We run the model twice: a first time without radial migration
(i.e. assuming independent annuli) and a second time with radial migration.

\begin{figure}
\begin{centering}
\includegraphics[width=0.5\textwidth]{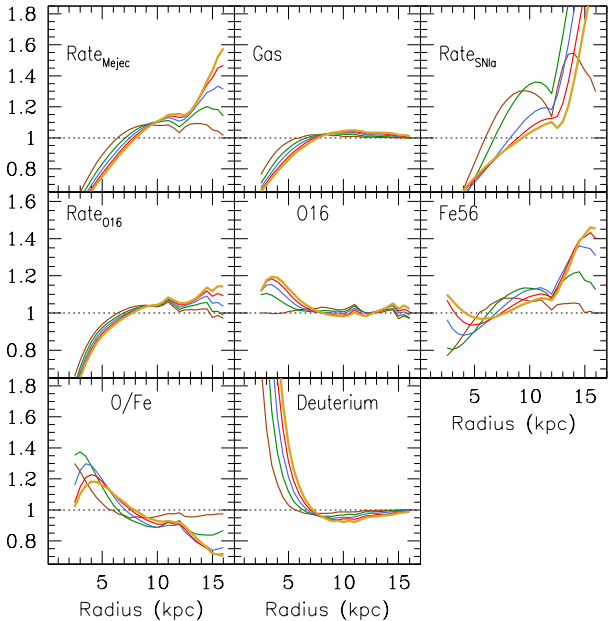} 
\par\end{centering}

\caption{\label{fig:migrvsNoMigr} Radial profiles of various quantities
obtained with radial migration \textit{divided by} the corresponding
profiles obtained without radial migration. Calculations are performed without IRA and results are displayed every two Gyr, the thickest curves corresponding to the final profiles at 10 Gyr. \textit{First row}: Rate
of mass ejection by stars, gas surface density and rate of SNIa. \textit{Second
row}: Rate of mass ejection of O-16 and gas abundance profiles of
O-16 and Fe-56. \textit{Third row}: Gas profiles of O/Fe ratio and deuterium. }
\end{figure}

The results of the two simulations are compared in Fig. \ref{fig:migrvsNoMigr},
which displays the final radial profiles of various quantities obtained
with radial migration \textit{divided by} the corresponding profiles
obtained without radial migration.

A first effect concerns the rate of mass ejection from long lived
sources (top right panel). Radial migration depopulates the inner
disk and populates the outer one with long lived (low mass) stars.
In the absence of radial migration the long lived stars return lately
a considerable amount of gas in the inner disk, where little gas is
left and ; with radial migration, a fraction of them does not return
that mass in their birth place, but in the outer disk, where they
migrate. However, the impact is not the same in the final gaseous
profile of the inner and outer disk (top middle panel): in the inner
disk, almost the totality of the gas is depleted early on from star
formation, and the gas non-returned from the migrated stars is a large
fraction of it; radial migration reduces the surface density of gas
in the inner disk. In contrast, in the outer disk, little of the initial
gas is consumed. The supplementary gas brought by the ejecta of migrated
stars barely changes the overall surface density there.

The rate of oxygen ejection (middle left panel) behaves in a similar way to the rate of
gas ejection (upper left panel), albeit for a non-intuitive reason: in the case of a
strong early star formation (as in this simulation), most of oxygen
is released at late times \textit{not by massive stars but by the
numerous intermediate and low-mass stars formed early on, which simply
release their initial oxygen} (their net yield being zero, there is
no chemical enrichment). In the inner regions, some of those stars
are missing because of radial migration, hence less oxygen is released
lately. The opposite holds for the outer regions. However, those considerations
do not impact on the oxygen profile, because only the massive stars
(which do not have time to migrate) enrich the local ISM with oxygen:
in the inner disk radial migration reduces the local gas amount (see
top middle panel and previous paragraph), and the dilution of the
same oxygen mass in a smaller gas amount results in a larger oxygen
abundance, by $\sim$20\%. The effect is negligible in the outer disk.

SNIa produce a large fraction of iron in a galaxy, from one to two
thirds, depending on the assumed prescription for their rate. In our
case, radial migration removes a large fraction of SNIa from the inner
disk and brings them in the outer disk (top right). The effect on
the Fe abundance profile is straightforward in the outer disk, where
Fe mass fraction is found to be $\sim$40\% larger with radial migration.
In the inner disk it is negligible: the missing Fe from migrated SNIa
is compensated in those regions by the effect of the Fe ejected from
massive star explosions being diluted in less gas (see previous paragraph
for oxygen).

The effects analyzed in the previous paragraphs are summarized in
the bottom left panel, displaying the O/Fe ratio: radial migration
of long-lived stars (including SNIa) makes the O/Fe ratio larger in
the inner disk (by up to 20\%) and smaller in the outer one (by 30\%).
Overall, it introduces a $\sim$50\% difference between the O/Fe ratios
at 5 kpc and 15 kpc; this corresponds to an increase of $\sim$0.02dex/kpc
in the radial gradient of {[}O/Fe{]} in the gaseous phase.

Finally, in the bottom middle panel we display an interesting effect
concerning the abundance of deuterium. Deuterium is produced only
in the Big Bang, it is only destroyed when passing in stellar interiors
(a process called astration) and most
of it is astrated in the numerous low and intermediate mass stars;
as a result, its abundance is steadily reduced in chemical evolution.
Radial migration removes a fraction of those stars from the inner
disk, hence the abundance of D in the gas of those regions is not
depleted as much as in the case of no migration but remains much higher,
by up to 80\%. In the outer disk, the D-free gas released by the migrators
has little effect, because there are considerable amounts of the initial
gas, which has not been consumed by star formation. We find then that,
overall, radial migration introduces an increase in the absolute value
of the D gradient of $\sim$0.025 dex/kpc.

\section{Summary\label{sec:Summary}}

In this work we study the effect of radial migration on the chemical evolution of a
bar-dominated disk galaxy, by analyzing a N-body+SPH simulation  and using appropriately tuned semi-analytical models.

We find that the non-dissipative fluid of stars behaves differently
from the dissipative fluid of gas regarding radial migration 
(Sec. 3): stars
experience strong radial migration over the whole disk, while gas
remains (on average) near its initial radius in the disk; however, in
the inner galaxy, the bar transfers large amounts of gas to the central
regions, affecting considerably the evolution of the developing bar/bulge.
Here we focus on radial migration of stars in the disk, leaving the
evolution of the bulge for a future study. By ''radial migration" we mean here
all types of radial displacement of stars, moving them away from their place of birth.

We studied  the impact of radial migration on various
aspects of the chemical evolution of the disk (Sec. 4), although our study was 
hampered by the use  of the Instantaneous
Recycling Approximation (IRA) and of a single metal (oxygen). 
 Radial migration increases
the stellar metallicity dispersion in all zones and all ages, 
but other factors contribute to that dispersion as well, namely the metallicity variations in the gas, due to incomplete mixing.
Because of the rapid evolution in all the model zones,
metallicity distributions differ little from one zone to another (their peak being close to the stellar yield), and radial migration does not change that result.
On the other hand, radial migration flattens the metallicity profiles,
bringing metal poor stars from the outer to the inner (metal-rich) zones and metal-rich stars from the inner to the outer (metal-poor) zones. Because
of that mixing, abundance profiles of old stellar populations of a given
age appear flatter from what they were at the period of stellar birth; as a consequence, their observation  cannot be used to infer the true evolution of the abundance profile.
The extent of that effect, however, depends a lot on the metallicity
profile of the disk: if it is relatively flat at the stars' birth, the effect
will be negligible, but if it is quite steep,  radial migration may not completely erase that signature. The same holds for azimuthal variations in the metallicities
of gas and stars. We argue that such variations are not a safe diagnostics
for radial migration, since they may have other origins (local infall of metal
poor gas or a metal poor merger). Finally, we obtain a U-shaped colour
profile, with old and red stars migrated from the inner to the outer disk
(as already found in Roskar et al. 2008).

Because of the limitations on chemical evolution imposed by the Instantaneous
Recycling Approximation adopted in the N-body+SPH simulation, we also
studied the effects of radial
migration on chemical evolution by adopting a different strategy.
We have shown (Sec. 5) that those effects
  can be studied to a good accuracy by post-processing
the results of a full N-body+SPH calculation with a simple chemical
evolution model having detailed chemistry and a parametrized description
of radial migration. We found that radial migration impacts on chemical
evolution both indirectly (by affecting the age-metallicity relations
and metallicity distributions of stars moved across the disk), and
directly, by moving around the long-lived nucleosynthesis sources (SNIa or AGB stars of 1.5 \ms)
and thus altering the abundance profiles of the gas; here (Sec. 5)
we have shown how the radial profiles of O, Fe and D are affected
in the case of our simulation, but other elements (like e.g. s-process
elements produced in AGB stars) may be concerned as well.
 Our 
post-processing results show clearly
that the full impact of radial migration on chemical evolution
cannot be evaluated with numerical codes using IRA.

It should be stressed, however, that the impact of radial migration
on chemical evolution depends on the system under study. Three factors have 
been identified up to now: strength of inhomegeneities in the gravitational potential, e.g. bar or spiral arms (stronger perturbations favoring
larger effects); duration of the radial migration (the longer
the bar acts, the larger the effects); and steepness of abundance profiles
at the time most of the stars are formed (steeper profiles favoring larger
effects). Those factors may cancel each other and mask the effects
of radial migration on chemical evolution: for instance, a strong bar driving
metal poor gas inwards 
will flatten the metallicity profile; the stars created
from that gas will display quasi-similar abundances all over the disk, even in the case of strong radial migration. 

This corresponds to the simulation studied here, although the reason
for the flat early metallicity profile is the evolution
 as a closed box with a rapid early star formation. 
In the case of a galaxy evolving more slowly inside-out, steeper
metallicity profiles are expected; however in that case, there is less time
left to an average star for radial migration, so its impact on chemical
evolution will not necessarily be more important.
Notice that continuous infall
of low metallicity gas, as expected in a galaxy like the Milky Way,
should also attenuate the impact of radial migration on gas chemistry,
by diluting radial abundance variations.

\medskip
\noindent
{\it Acknowledgments} E.A. acknowledges financial
support by the CNES and by the European Commission through the
DAGAL Network (PITN-GA-2011-289313). We are grateful to the referee for his/her
thorough and constructive report.

\bibliography{reference}

\end{document}